\pdfoutput=1

\documentclass[iop]{emulateapj}

\usepackage[svgnames,pdftex,hyperref]{xcolor}
\usepackage[pdftex]{hyperref}
\hypersetup{breaklinks=True,colorlinks=True,linkcolor=Black,citecolor=Black,
urlcolor=Black,pagebackref,pdftitle={PSR J2129-0429},pdfauthor={Eric Bellm}}
\usepackage{nicefrac}

\usepackage{xspace}

\usepackage{ amssymb }
\usepackage{amsmath} 
\usepackage{aas_macros}

\newcommand{\psr}{PSR\,J2129$-$0429\xspace}
\newcommand{\angstrom}{\textup{\AA}\xspace}
\newcommand{\msun}{\ensuremath{M_\odot}\xspace}

\newcommand{\ppm}{\ensuremath{\pm}}

\slugcomment{Submitted to ApJ}

\shorttitle{\psr}
\shortauthors{Bellm et al.}

\begin{document}



\title{Properties and Evolution of the Redback Millisecond Pulsar Binary \psr}


\author{Eric C. Bellm\altaffilmark{1,2},
David L. Kaplan\altaffilmark{3},
Rene P. Breton\altaffilmark{4,5},
E. Sterl Phinney\altaffilmark{1},
Varun B. Bhalerao\altaffilmark{6},
Fernando Camilo\altaffilmark{7},
Sumit Dahal\altaffilmark{8},
S. G. Djorgovski\altaffilmark{1},
Andrew J. Drake\altaffilmark{1},
J. W. T. Hessels\altaffilmark{9,10},
Russ R. Laher\altaffilmark{11},
David B. Levitan\altaffilmark{1},
Fraser Lewis\altaffilmark{12,13},
Ashish A. Mahabal\altaffilmark{1},
Eran O. Ofek\altaffilmark{14},
Thomas A. Prince\altaffilmark{1},
Scott M. Ransom\altaffilmark{15},
Mallory S. E. Roberts\altaffilmark{16},
David M. Russell\altaffilmark{8},
Branimir Sesar\altaffilmark{17},
Jason A. Surace\altaffilmark{11},
Sumin Tang\altaffilmark{1}
}

\altaffiltext{1}{Cahill Center for Astronomy and Astrophysics, California
Institute of Technology, Pasadena, CA 91125}
\altaffiltext{2}{ebellm@caltech.edu}
\altaffiltext{3}{Department of Physics, University of Wisconsin-Milwaukee,
Milwaukee, WI 53201, USA}
\altaffiltext{4}{Jodrell Bank Centre for Astrophysics, Alan Turing
Building, School of Physics and Astronomy, The University of Manchester,
Oxford Road, Manchester, M13 9PL, U.K.}
\altaffiltext{5}{School of Physics and Astronomy, University of
Southampton, Southampton, SO17 1BJ, U.K.}
\altaffiltext{6}{Inter University Centre for Astronomy and Astrophysics, PO
Bag 4, Ganeshkhind, Pune 411007, India}
\altaffiltext{7}{Columbia Astrophysics Laboratory, Columbia University, New York, NY 10027, USA}
\altaffiltext{8}{New York University Abu Dhabi, PO Box 129188, Abu Dhabi,
UAE}
\altaffiltext{9}{ASTRON, the Netherlands Institute for Radio Astronomy,
Postbus 2, 7990 AA, Dwingeloo, The Netherlands}
\altaffiltext{10}{Anton Pannekoek Institute for Astronomy, University of
Amsterdam, Science Park 904, 1098 XH Amsterdam, The Netherlands}
\altaffiltext{11}{Infrared Processing and Analysis Center, 
California Institute of Technology, Pasadena, CA 91125}
\altaffiltext{12}{Faulkes Telescope Project, School of Physics and
Astronomy, Cardiff University, 5 The Parade, Cardiff, CF24 3AA, Wales, UK}
\altaffiltext{13}{Astrophysics Research Institute, Liverpool John Moores
University, IC2, Liverpool Science Park, 146 Brownlow Hill, Liverpool L3
5RF, UK}
\altaffiltext{14}{Benoziyo Center for Astrophysics, Faculty of Physics,
Weizmann Institute of Science, Rehovot 76100, Israel}
\altaffiltext{15}{National Radio Astronomy Observatory, 520 Edgemont Road,
Charlottesville, Virginia 22903-2475, USA}
\altaffiltext{16}{Eureka Scientific Inc., 2452 Delmer Street, Suite 100,
Oakland, California 94602-3017, USA}
\altaffiltext{17}{Max Planck Institute for Astronomy, K\"onigstuhl 17, D-69117
Heidelberg, Germany}







\begin{abstract}

\psr is a ``redback'' eclipsing millisecond pulsar binary with an unusually
long 15.2\,hour orbit. It was discovered by the Green Bank Telescope in a
targeted search of unidentified Fermi gamma-ray sources.  
The pulsar companion is optically bright (mean $m_R = 16.6$\,mag), allowing
us to construct the longest baseline photometric dataset available 
for such a system.
We present ten years of archival and new photometry of the 
companion from LINEAR, CRTS, PTF, the Palomar 60-inch, and LCOGT.  
Radial
velocity spectroscopy using the Double-Beam Spectrograph 
on the Palomar 200-inch indicates that the
pulsar is massive: $1.74\pm0.18$\,\msun.  The G-type pulsar companion
has mass $0.44\pm0.04$\,\msun, one of the heaviest known redback companions.
It is currently 95$\pm1$\% Roche-lobe filling and only mildly irradiated by the
pulsar.  
We identify a clear 
13.1\,mmag\,yr$^{-1}$ secular decline in the mean magnitude of the
companion as well as smaller-scale variations in the optical lightcurve
shape.  This behavior may indicate that the companion is cooling.
Binary evolution calculations indicate that \psr has an orbital period
almost exactly at the bifurcation period between systems that converge into
tighter orbits as black widows and redbacks and those that diverge into
wider pulsar--white dwarf binaries.  Its eventual fate may depend on whether it
undergoes future episodes of mass transfer and increased irradiation.

\end{abstract}




\section{Introduction}

Millisecond pulsars (MSPs) eclipsed by material from their low-mass binary
companions may provide a major path to the production of isolated MSPs:
after spinning up the pulsar through accretion \citep{Alpar:82:Recycling,
Radhakrishnan:82:Recycling, Bhattacharya:91:MSPEvolution}, the pulsar wind
may ablate away the donor
\citep{vandenHeuvel:88:Ablation,Phinney:88:BlackWidowModel,
Kluzniak:88:1957Ablation}.  This cannibalism suggested the name ``Black
Widow'' upon the discovery of the first example, which had a companion mass
of only a few percent \msun \citep{Fruchter:88:B1957Discovery}.  Later
observations and theoretical work suggested that the ablation rate for the
first Black Widow system might be too slow to evaporate its companion in a
Hubble time, however
\citep{Levinson:91:PulsarAblation,Fruchter:92:BlackWidowEclipse}. 
Ablation may be more efficient in other systems \citep[e.g.,][]{Bailes:11:MSPPlanet},
such that the companion is completely destroyed or only a small remnant 
remains.
Instabilities in very low mass companions \citep{Deloye:03:UCBDisruption}
or dynamical encounters in globular clusters
\citep{King:03:BWDynamicEjection} provide
alternative means to produce isolated MSPs.

Today, thanks to targeted radio and X-ray surveys
of globular clusters and, more recently, in regions of the 
broader Galactic Field where
\textit{Fermi}-LAT has localized sources,
more than 30 black widow systems are known \citep[][and references
therein]{Roberts:13:Spiders}.  Moreover, many ``redback''
systems\footnote{Redback spiders are the Australian cousins of Black Widow
spiders.} with
higher-mass, non-degenerate companions of 0.2--0.7 \msun have been
discovered (\citealp{Roberts:13:Spiders}, and references therein;
\citealp{Crawford:13:J1723Mass}).  
The discovery of several redback systems that transition
between accretion-powered low-mass X-ray binary states and rotation-powered radio
pulsar states has provided further support for the
recycling scenario
\citep{Archibald:09:PulsarBinaryLink,Papitto:13:J1824Transition,Bassa:14:XSSJ12270Transition,Stappers:14:J1023StateChange,Patruno:14:J1023Accretion}.
Irradiation feedback and accretion disk instabilities 
may be responsible for the relatively short timescale
transitions (years) observed between accreting and detached states
\citep{Benvenuto:15:RedbackQuasiRLOF}.
Current models disagree as to whether redbacks and black widows are
distinct populations that evolve separately due to bimodal evaporation
efficiency \citep{Chen:13:RBBWEvolution} 
or whether black widow systems are
an endpoint in the evolution of some redback systems
\citep{Benvenuto:14:RBBWEvolution}.

In the optical bands, black widow and redback binaries show large-amplitude
variability due to effects such as the ellipsoidal modulation of the
near-Roche filling companion, dayside and nightside temperature
differentials, and gravity darkening.  The
apparent magnitudes of these systems vary considerably with distance and
the intrinsic luminosity of their companions.  
With their larger companions
and relative proximity, redbacks in the Galactic Field 
are the brightest of these systems in the optical band, 
allowing for detailed photometric and
spectroscopic monitoring with moderate aperture telescopes
\citep[e.g.,][]{Breton:13:OpticalBW,Li:14:FermiCounterparts,
Schroeder:14:LCModelsNSMass}.
MSP binaries in the Field are also useful tracers of binary evolution,
as unlike systems in globular clusters they are unlikely to experience
dynamical exchange encounters.

With a radio timing solution and an optical radial velocity amplitude from
the companion, it is possible to measure the mass ratio of the pulsar and
its companion.  While expected theoretically
\citep[e.g.,][]{Phinney:94:BinaryMSPs,Benvenuto:14:RBBWEvolution}, the
neutron star masses in the black widow and redback systems measured to date
have largely  
proven to be heavier than 1.4 \msun
\citep{vanKerkwijk:11:B1957Mass, Romani:12:J1311HeavyNS,
Deller:12:J1023Mass, Crawford:13:J1723Mass}.  
Measurement of 
pulsar masses in these systems 
thus provides a valuable expansion of the sample needed to
understand the recycling process.  While the most precise 
measurements of the masses of high-mass neutron
stars have been obtained through measurement of Shapiro delay
\citep{Nice:05:MassiveNS,Demorest:10:Shapiro,Freire:11:Shapiro}, discovery
of additional heavy neutron stars in pulsar binaries of various types
\citep[e.g.,][]{Bassa:06:MassiveNS,Kaplan:13:J1816,Antoniadis:13:MassiveNS,
tmp_Strader:15:MassiveNS} can also inform studies of the equation of state
of neutron stars \citep[for a review, see][]{Lattimer:12:NSEOSReview}.

\psr is a redback system with a bright (mean $m_R = 16.6$) optical
counterpart.
The gamma-ray counterpart to \psr appeared in the first \textit{Fermi}
source catalog as 1FGL\,J2129.8$-$0427 \citep{Abdo:10:1fgl} and in subsequent
versions as 2FGL J2129.8$-$0428 \citep{Nolan:12:2fgl} and 3FGL\,J2129.6$-$0427
\citep{tmp_LAT:15:3fgl}.  
The \textit{Fermi} spectrum does not show significant curvature, making it
difficult to firmly classify as a pulsar based on the gamma-ray data alone
\citep[e.g.,][]{Ackermann:12:FermiClassification}.

Nevertheless, a radio survey of the 1FGL error box with the Green Bank
Telescope identified a pulsar counterpart with a spin period of 7.62\,ms
\citep{Hessels:11:GBTFermiSurvey}.  The dispersion measure of
16.9\,pc\,cm$^{-3}$ implies a distance of 0.9\,kpc using the NE2001 model
\citep{Cordes:02:NE2001}.
Phase-folded X-ray observations of the system show an unusual double-peaked
phase profile from the intra-binary shock, suggesting a compact
but structured emission region
\citep{Roberts:15:RedbacksXrays}.

In this paper we present optical photometry and spectroscopy of 
the \psr binary system. 
In \S \ref{sec:observations} we describe the
observations.  We perform fits to the radial velocity amplitude in \S
\ref{sec:rvs} and to the system geometry in \S \ref{sec:binary}.  In \S
\ref{sec:longterm} we present evidence for secular evolution in the
lightcurve.  We close with a discussion of the evolution of \psr 
(\S \ref{sec:discussion}).

Throughout the paper, we use the convention that the zero point of
orbital phase occurs when the companion is between the pulsar and the
Earth (companion inferior conjunction).  
The Time of Ascending Node is accordingly at phase 0.75 in this convention.

\section{Observations} \label{sec:observations}

\subsection{Photometry}

Using the localization of \psr provided by a preliminary radio-derived pulsar
timing ephemeris (21h29m45.039$\pm$0.001s, $-04^\circ29^\prime05.59\pm0.08^{\prime\prime}$
(J2000); Bangale et al., in prep), we
found a bright ($m_R \sim 16.5$) 
optical counterpart in the databases of several 
time-domain surveys that varied at the orbital period of 15.245\,hr.
We also acquired new multicolor phase-resolved observations to enable
detailed lightcurve modeling.

The optical counterpart of \psr was repeatedly detected in imaging
conducted by the Palomar Transient Factory (PTF)
\citep{Law:09:PTFOverview,Rau:09:PTFScience} on the Palomar Samuel Oschin
48-inch Schmidt (P48) at 21h29m45.06s,
$-04^\circ29^\prime06.8^{\prime\prime}$ (J2000)\footnote{Typical PTF
absolute astrometric errors are 0.1--0.2$^{\prime\prime}$ RMS. The
1.25$^{\prime\prime}$ distance of the radio position may be due to
covariances between the orbital derivatives and spindown in the relatively
short radio timing baseline. The SDSS DR12 \citep{Alam:15:SDSSDR12}
position of the object (21h29m45.05s,
$-04^\circ29^\prime06.83^{\prime\prime}$) is consistent with the PTF
position.}. We obtained calibrated aperture photometry for \psr from the
standard PTF photometric pipeline, which includes image reduction
\citep{Laher:14:PTFPipeline}, photometric calibration
\citep{Ofek:12:PhotometricCal} to the Sloan Digital Sky Survey
\citep{York:00:SDSS}, and a relative photometry correction for each image
to remove any residual scalar offsets in the zero-points due to
non-photometric conditions \citep[c.f.][Appendix
A]{Ofek:11:RelativePhotometry}.  We added a systematic error of 0.01\,mag
in quadrature with the pipeline-generated photometric errors so that a
constant fit to a nearby ($53.6^{\prime\prime}$ separation) comparison star
(21h29m46.9s, $-04^\circ28^\prime20.5^{\prime\prime}$, $m_R = 16.6$\,mag)
had reduced chi-squared $\chi^2_\nu = 1$.

\psr is also present in data from the Catalina Real-time Transient Survey
\citep[CRTS;][]{Drake:09:CRTS} catalog and in its associated catalog of
periodic sources \citep{Drake:14:PeriodicSources}.  
The Catalina images are spaced ten minutes
apart to look for asteroid motion. The magnitudes are obtained by running
SExtractor \citep{Bertin:96:SExtractor} on the images and then calibrating
them to the V band\footnote{
More details for the CRTS Second Data Release (DR2) magnitudes can
be found at
\url{http://nesssi.cacr.caltech.edu/DataRelease/FAQ2.html\#calib}. The set
of observations used here included dates outside the DR2 release. 
Magnitudes from these observations are taken from the transient pipeline
and lack the small frame offsets applied in DR2.},
as the raw CCD images are unfiltered. 

Additionally, the system was imaged by the Lincoln Near-Earth Asteroid
Research survey \citep[LINEAR;][]{Stokes:2000:LINEAR, Sesar:11:LINEAR}.  
We performed PSF photometry on the images and
photometrically calibrated the images to SDSS.

We acquired targeted 
phase-resolved observations of \psr in $g^\prime$, $r^\prime$,
and $i^\prime$ using the CCD camera on the 
roboticized Palomar 60-inch telescope \citep[P60;][]{Cenko:06:P60}
between November 2011 and June 2012. Basic
image reductions were performed by the automated pipeline.  We obtained
aperture photometry for the source and companion stars using SExtractor
and calibrated the photometry to SDSS.

We obtained additional $g^\prime$, $r^\prime$, $i^\prime$ and $z^\prime$
imaging of \psr using the Las Cumbres Observatory Global Telescope
(LCOGT) Network over a period of a month starting 26 October 2014. 
Observations were made using the 1-m LCOGT network sites at Siding Spring
Observatory, Australia, Cerro Tololo Inter-American Observatory, Chile,
McDonald Observatory, USA and SAAO, South Africa, as well as the 2-m
Faulkes Telescopes at Haleakala Observatory, Maui, Hawaii, USA and Siding
Spring Observatory, Australia.
The data
reduction was performed using LCOGT's automatic pipeline and aperture
photometry was conducted with IRAF version 2.16.1
\citep{Tody:86:IRAF} and calibrated to SDSS using four bright,
unsaturated stars visible in all images.

All times were corrected to the Solar System barycenter.
The photometric observations are summarized in Table~\ref{tab:phot}.

\begin{deluxetable*}{l c l l l r r}
  \tablewidth{0pt}
\tablecaption{Summary of Photometric Observations\label{tab:phot}}
  \tablehead{
\colhead{Telescope} & \colhead{Filter} & \colhead{Start Date} & \colhead{End Date}
& \colhead{Exposure} & \colhead{Number} &
\colhead{Precision} \\
 & & & & \colhead{(s)} & & \colhead{(mmag)}
}
\startdata
LINEAR & open &  2003~May~30 & 2011~Oct.~29 & 3--18 & 470 & 30 \\
CRTS & open & 2005~May~12 & 2014~May~5 &  30 & 365 & 76 \\
P48 & Mould $R$ & 2009~Jun.~26 & 2014~Oct.~11 & 60 & 116 & 14 \\
P60 & $g^\prime$ & 2011~Nov.~10 & 2012~Jun.~14 & 60, 120 & 41 & 26 \\
P60 & $r^\prime$ & 2011~Nov.~10 & 2012~Jun.~14 & 60, 120 & 43 & 20 \\
P60 & $i^\prime$ & 2011~Nov.~10 & 2012~Jun.~14 & 60, 120 & 50 & 18 \\  
LCOGT 1 \& 2\,m & $g^\prime$ & 2014~Oct.~26 & 2014~Nov.~22 & 100--200 & 25 & 29
\\
LCOGT 1 \& 2\,m & $r^\prime$ & 2014~Oct.~26 & 2014~Nov.~22 & 100--200 & 25 & 20
\\
LCOGT 1 \& 2\,m & $i^\prime$ & 2014~Oct.~26 & 2014~Nov.~22 & 100--200 & 23 & 28
\\
LCOGT 1 \& 2\,m & Pan-STARRS $Z$ & 2014~Oct.~26 & 2014~Nov.~22 & 100--200 & 13 & 69 
\enddata
\end{deluxetable*}

\subsection{Spectroscopy} \label{sec:spectra}
\begin{figure*}
\includegraphics[width=\textwidth]{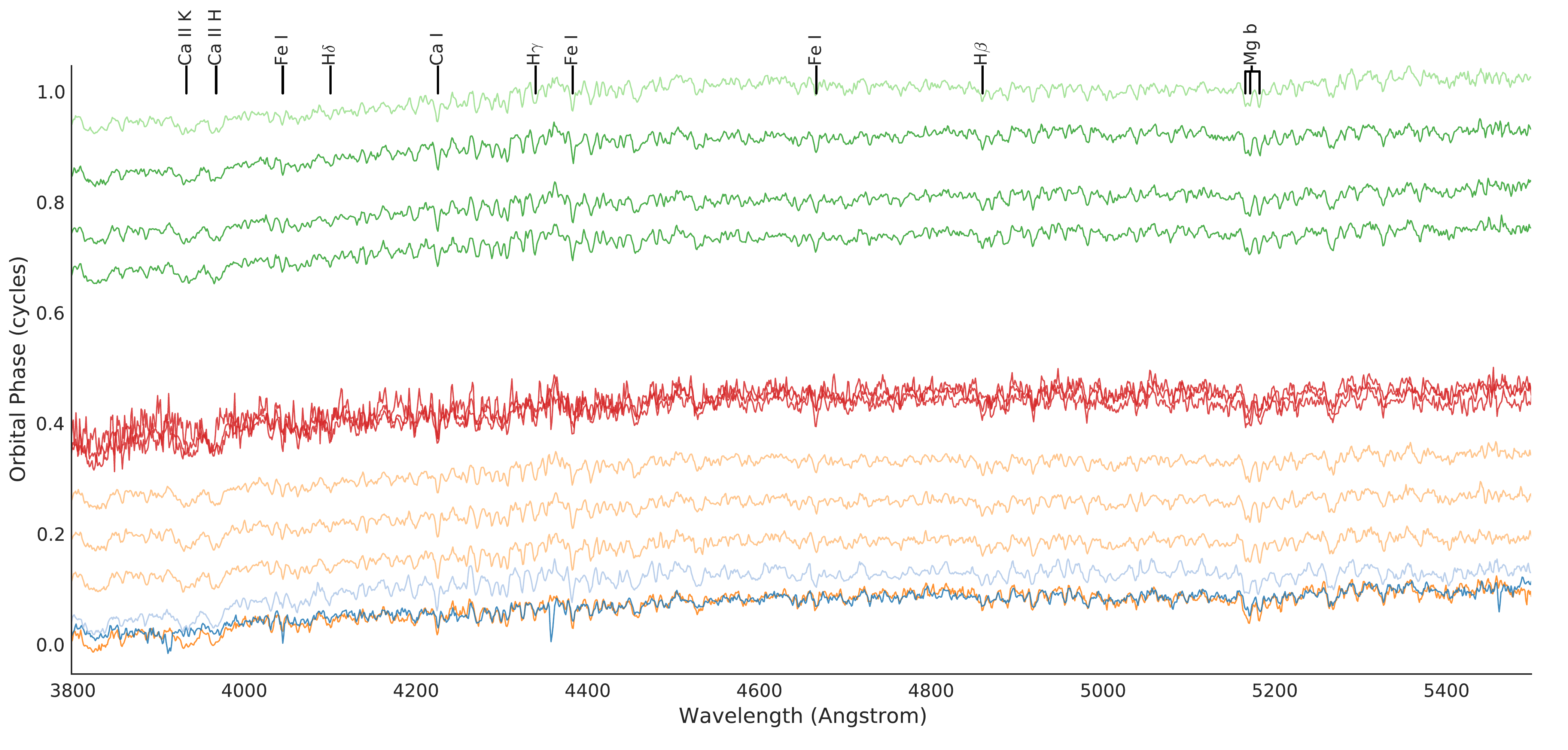}
\caption{Blue-arm DBSP spectra of the \psr companion
ordered by pulsar orbital phase.  Spectra are adjusted to zero velocity.
Spectra obtained within one night of observing are plotted in the same
color; Table \ref{tab:spec} lists the orbital phase of each spectrum. 
\label{fig:blue_spec}}
\end{figure*}

\begin{figure*}
\includegraphics[width=\textwidth]{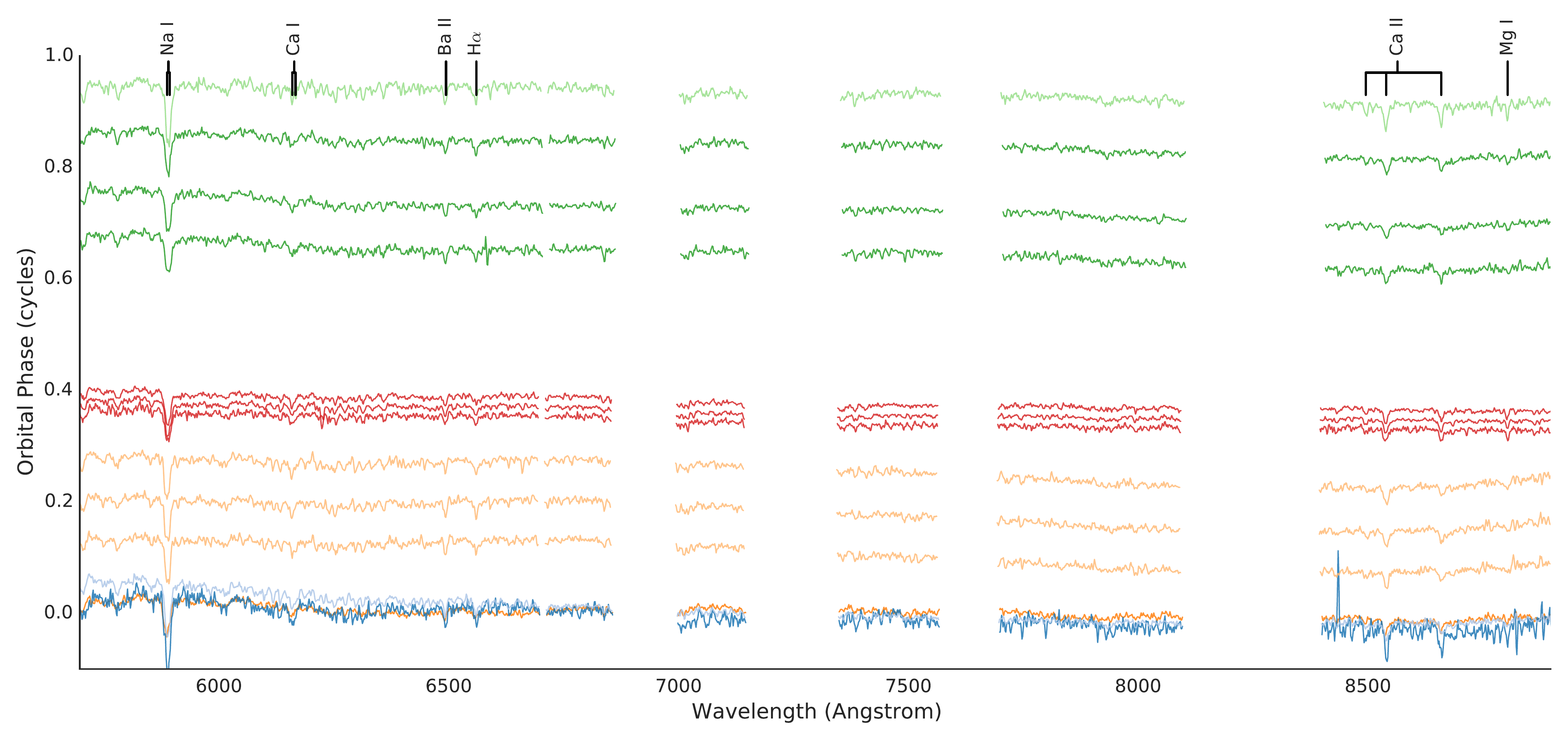}
\caption{Red-arm DBSP spectra of the \psr companion
ordered by pulsar orbital phase.  Colors are as in Figure
\ref{fig:blue_spec}.  Regions of telluric absorption are masked, as is a
small region near 6707\,\AA\ affected by a dome lamp fluorescence feature.
\label{fig:red_spec}}
\end{figure*}

We obtained spectra of \psr using the Palomar 200-inch Hale telescope and
the Double-Beam Spectrograph \citep[DBSP;][]{Oke:82:DBSP} using the new red
camera \citep{Rahmer:12:DBSPRed}.
Table \ref{tab:spec} lists the observation parameters.
We reduced data from both arms of the spectrograph using a custom PyRAF-based
pipeline\footnote{\url{https://github.com/ebellm/pyraf-dbsp}}.  
The pipeline performs standard image processing and spectral reduction
procedures, including bias subtraction, flat-field correction, wavelength
calibration, optimal spectral extraction, and flux calibration.  To account
for instrument flexure in individual science spectra, we subtract an average 
wavelength offset computed from known sky lines on a per-frame basis
\citep{Sesar:13:OrphanStream}.  The
RMS scatter of the corrected sky line positions provides an estimate of the
uncertainty in the wavelength solution, which we add in quadrature with the
template fitting uncertainty when computing radial velocities (Section
\ref{sec:rvs}).  The mean uncertainty for these observations was
6.2\,km\,s$^{-1}$ at 4750\,\angstrom on the blue side and 8.6\,km\,s$^{-1}$
at 7400\,\angstrom on the red side.  Figures \ref{fig:blue_spec} and
\ref{fig:red_spec} show the reduced spectra ordered by pulsar orbital
phase.

The spectra indicate that the companion of \psr is non-degenerate, with
features resembling a G-type star.  Some variation in the line
strengths is observed. From the few observations we have that are repeated 
near phase 0, there appears to be stochastic variability in
addition to phase-dependent differences (Figures \ref{fig:blue_spec}
and \ref{fig:red_spec}).

\begin{deluxetable*}{l c c c c}
  \tablewidth{0pt}
\tablecaption{Summary of Spectroscopic Observations.  \label{tab:spec}}
  \tablehead{
\colhead{Date (UTC)} & \colhead{Epoch at Barycenter} & \colhead{Orbital Phase} 
& \colhead{Aperture} & \colhead{Exposure Time} \\
& \colhead{(TDB MJD)} & & \colhead{(arcsec)} & \colhead{(s)} 
}
\startdata
2011-12-27 &  55922.07355 & 0.02 &  1.0 &  900 \\
2012-05-29 &  56076.45451 & 0.06 &  1.0 &  900 \\
2012-07-15 &  56123.43948 & 0.02 &  1.0 &  600 \\
2012-09-20 &  56190.20842 & 0.13 &  0.5 &  900 \\
2012-09-20 &  56190.25525 & 0.21 &  0.5 &  900 \\
2012-09-20 &  56190.30287 & 0.28 &  0.5 &  900 \\
2012-09-21 &  56191.19072 & 0.68 &  0.5 &  500 \\
2012-09-21 &  56191.24078 & 0.76 &  0.5 &  900 \\
2012-09-21 &  56191.30869 & 0.86 &  0.5 &  900 \\
2013-06-02 &  56445.45257 & 0.95 &  0.5 &  900 \\
2013-07-04 &  56477.47904 & 0.36 &  1.5 &  900 \\
2013-07-04 &  56477.48975 & 0.38 &  1.5 &  900 \\
2013-07-04 &  56477.50047 & 0.40 &  1.5 &  900 
\enddata
\tablecomments{All observations were
conducted with the Double Beam Spectrograph of the Palomar 200-inch Hale
Telescope.  Gratings used were 600 lines/mm blazed at 4000 \angstrom (blue
side) and 316 lines/mm blazed at 7500 \angstrom (red side) for all dates
except 2013 July 4, when the 1200/5000 and 1200/7100 gratings were used.}
\end{deluxetable*}

\section{Radial Velocity} \label{sec:rvs}
\begin{figure}
\includegraphics[width=\columnwidth]{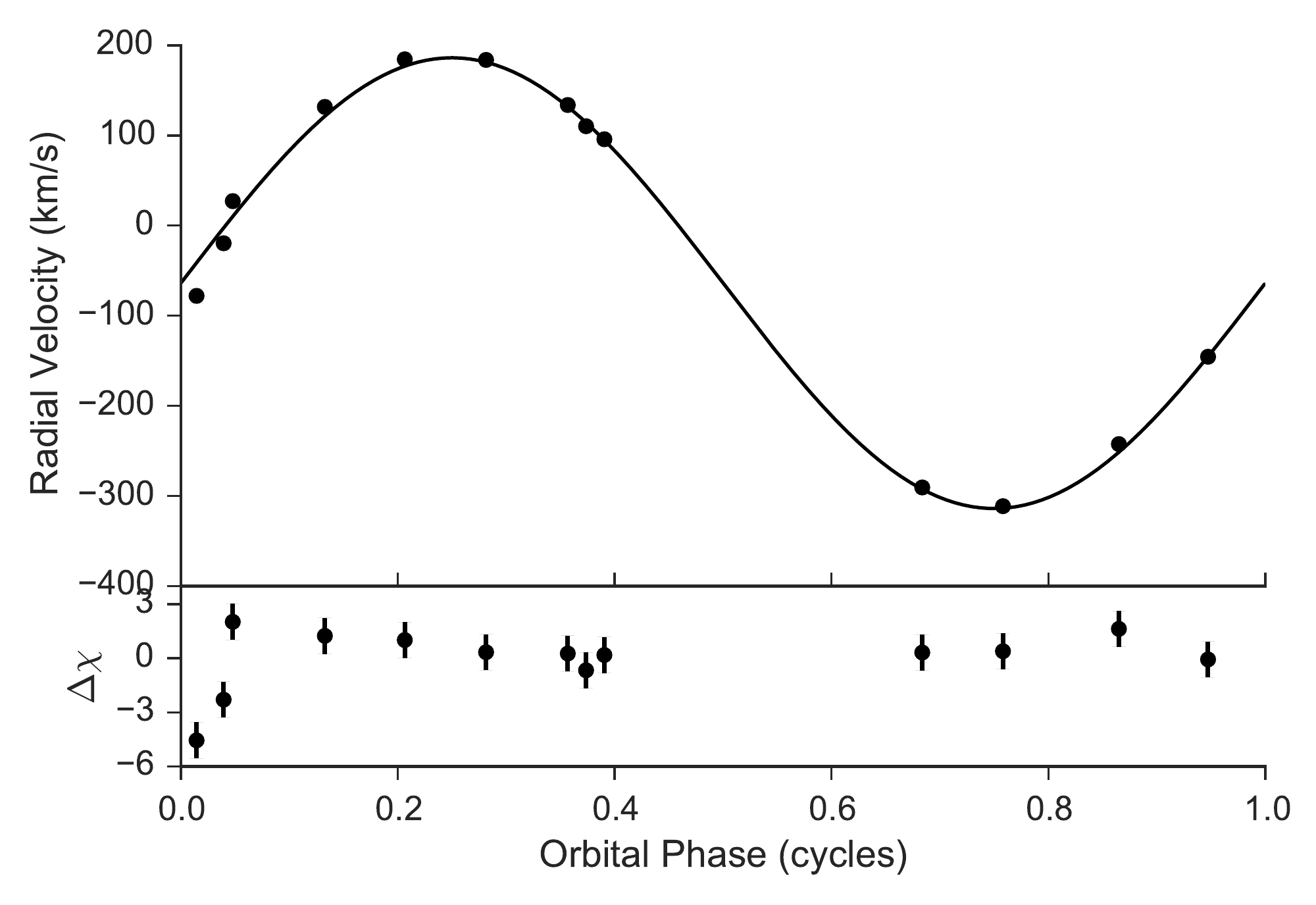}
\caption{Radial velocity values, best-fit curve, and sigma residuals 
for \psr from the DBSP red arm data.  Error bars are comparable to the
symbol size.
The best fit amplitude is $K=250.0\pm3.7$\,km\,s$^{-1}$ and the 
systemic velocity is $\gamma = -63.9\pm 2.4$\,km\,s$^{-1}$.
\label{fig:rvs}}
\end{figure}

We obtained radial velocity measurements from our spectra using a
template-fitting approach as implemented in \citet{Bhalerao:12:HMXBRV}.  We
fit flux-calibrated spectra with the stellar atmosphere models of
\citet{Munari:05:Atmospheres}, using a linear polynomial to account for
any residual fluxing errors.  We included an extinction coefficient of  
$A_V =  0.10$\,mag \citep{Schlafly:11:Extinction}.  We fit the red-arm
data from 5700--8900\,\angstrom, omitting telluric bands at
6860--7000\,\angstrom, 7570--7700\,\angstrom, 7150--7350\,\angstrom, and
8100--8400\,\angstrom.  The best-fit atmosphere model at all phases 
had $T=5750$\,K, $\log g = 5.0$, solar metallicity, and 100\,km\,s$^{-1}$ 
rotational broadening.  (These values are broadly consistent with the
values we derive from binary modeling in Section \ref{sec:binary} and our
assumption of corotation.)  By
stepping the model through a grid of wavelength offsets, we determined the 
radial velocity at each epoch.  A quadratic fit to the $\chi^2$ surface
provided $1\sigma$ error estimates, which we added in quadrature with the
wavelength calibration uncertainties obtained in Section \ref{sec:spectra}.
Finally, we converted the observed radial velocities to the Solar System
barycenter using Python routines 
derived from the Standards of Fundamental Astronomy
Library \citep{citeSOFA}.
The $\chi^2$/dof for the fit is 11.6/11.
Figure \ref{fig:rvs} shows the resulting radial velocities and best-fit
radial velocity amplitude using the radio ephemeris ($P_{\rm orb} =
0.63522741310 \pm 3.3\times10^{-10}$\,days;  
TASC $= 55702.111161463 \pm  9.8\times 10^{-8}$)\footnote{We do not include
radio-derived orbital period derivatives because 
the time baseline for our optical data
extends well beyond the radio timing interval, which would lead to
significant extrapolation errors from the higher-order period derivatives.}
We find a
radial velocity amplitude $K=250.0\pm3.7\,{\rm km\,s}^{-1}$.  Together
with the measured projected semi-major axis (1.85\,lt-sec) from pulsar timing we
find a mass ratio $q=3.93\pm0.06$.  
Because the irradiation of the
companion by the pulsar is modest and the mass ratio less extreme than in
black widow systems \citep[e.g.,][]{vanKerkwijk:11:B1957Mass}, the correction of the measured center of light 
RV amplitude to the center of mass is small ($\sim0.2$\%), removing a
potential source of systematic error.  We nevertheless model this
correction self-consistently in our binary fitting in the next section.

\section{Binary fitting} \label{sec:binary}

We fit the available multicolor photometry from P60, and LCOGT 
as well as the 2009--2010 PTF data to a model of a
distorted, irradiated binary companion using \texttt{Icarus}
\citep{Breton:13:OpticalBW}.  This fitting computed
filter-dependent photometry for a companion including the effects of
irradiation and ellipsoidal modulation, which enable us to constrain
the inclination, the masses of the binary members, the distance, the
degree of irradiation, etc.  Our fitting made use of prior probability
distributions on the distance, extinction, and radial velocity
amplitude.  The distance prior was based on the dispersion measure
(DM) distance of 900\,pc \citep{Cordes:02:NE2001} with a
conservative estimated uncertainty of 50\% ($\pm 450\,$pc).  The
radial velocity amplitude prior was based on the measured radial
velocity amplitude as well as the pulsar's projected semi-major axis,
and we used $K=250\pm4\,{\rm km\,s}^{-1}$ as determined above.
Finally we determined a prior on the extinction $A_V$ based on the
measured X-ray column density $N_{\rm H}$ 
and the observed $A_V(N_{\rm
  H})$ relation:
\begin{equation}
  p(A_V) = 0.65 {\cal N}(0.131,\,0.082) + 0.35{\cal
    N}(0.068,\,0.188)
  \label{eqn:AV}
\end{equation}
(Roberts et al., in prep).
Here, ${\cal N}(\mu,\sigma)$ represents a Gaussian probability
distribution function (pdf) with mean $\mu$ and standard deviation $\sigma$;
both pdfs were properly truncated so that $A_V \geq 0$.
Otherwise, we assumed flat priors on $\cos i$ (where $i$ is the binary
inclination), the nightside temperature $T_{\rm night}$, the dayside
temperature $T_{\rm day}$, and the Roche-lobe filling factor $f$ (here $f=R_{\rm
  nose}/R_{\rm L1}$ is the ratio of the radius of the companion pointing
to the pulsar to that of the L1 point).

\begin{figure*}
  \plotone{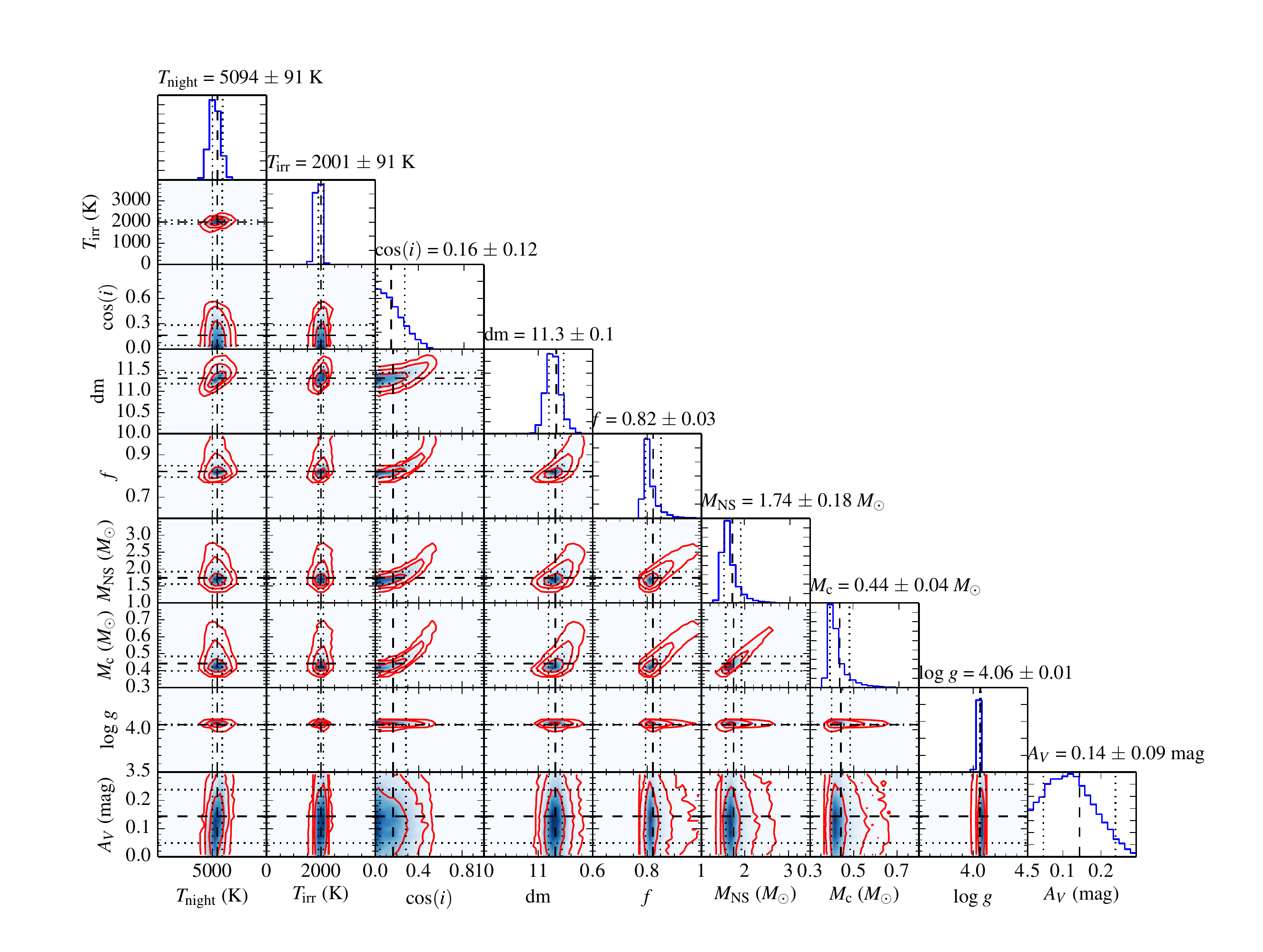}
  \caption{Joint two-dimensional posterior probability functions from
    the MCMC analysis.  We show 1-, 2-, and 3-$\sigma$ joint
    probability contours for each pair of parameters as well as the
    marginalized one-dimensional posterior pdfs for each parameter.
    See Table~\ref{tab:fit} for details.}
  \label{fig:fit}
\end{figure*}

\begin{figure}
\includegraphics[width=\columnwidth]{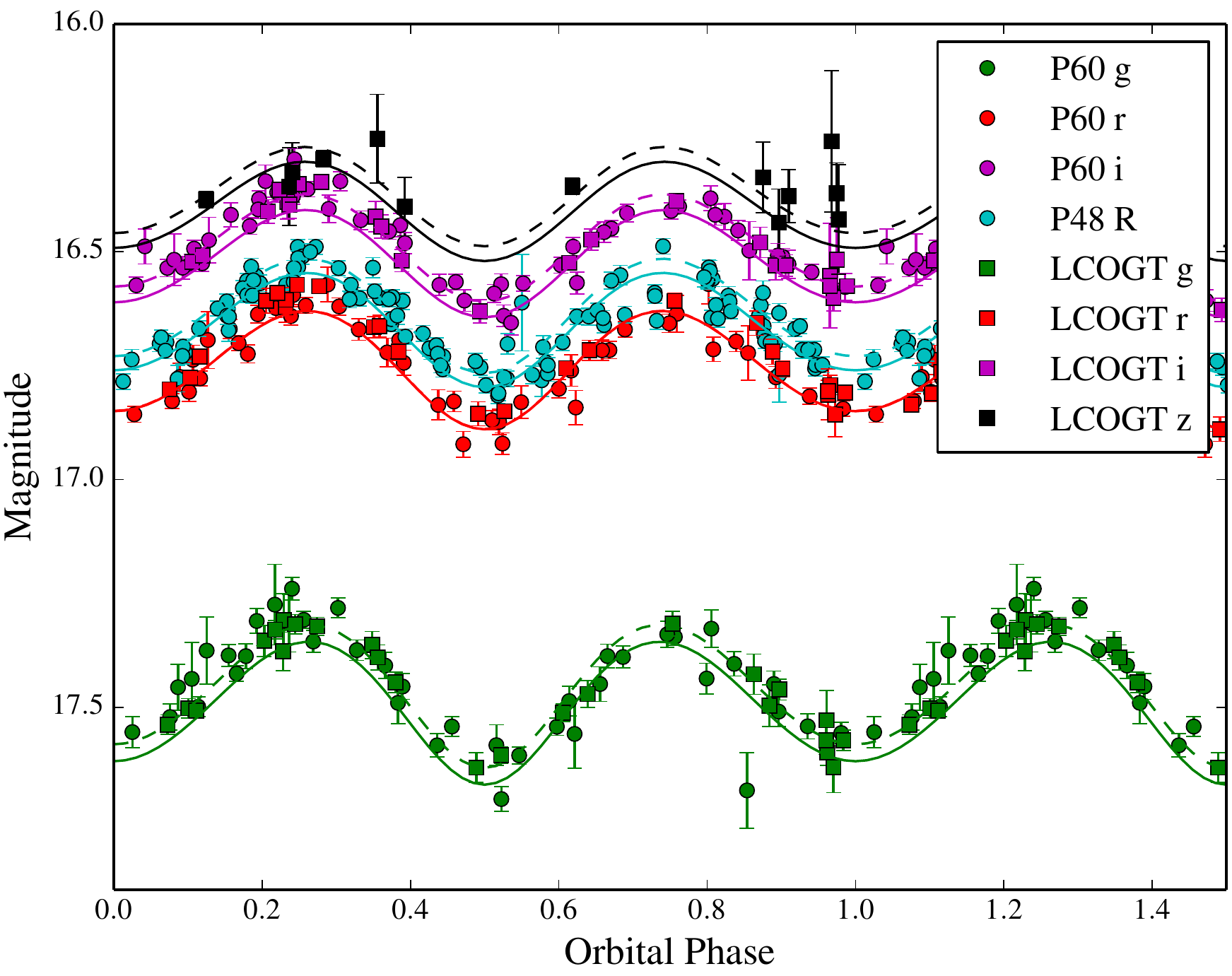}
\caption{Best-fit lightcurve for \psr, using the P60, P48, and LCOGT
  photometry.  The lightcurve is repeated 1.5 times for clarity.  For
  each band we show the nominal model lightcurve (solid line) as well
  as the lightcurve with shifts to account for zero-point uncertainties
  (dashed line), all of which were $<0.04\,$mag.    
  }
\label{fig:lightcurve}
\end{figure}

We performed a Markov-Chain Monte Carlo (MCMC) analysis of the
photometry, fitting for $\cos i$, $T_{\rm night}$, $T_{\rm day}$, $K$, ${\rm
DM}$, $A_V$, and $f$, where ${\rm DM}=-2.5\log_{10}(d/10\,{\rm pc})$ is the distance
modulus. We included a 0.01\,mag systematic uncertainty on each
measurement, and we also allowed for individual offsets for each
telescope/filter combination with an assumed zero-point uncertainty of
0.05\,mag\footnote{These normalization offsets remove the largest observed 
source of secular variation (\S\ref{sec:longterm}).  The fits are dominated
by the P60 data, taken in 2011--2012.  Little color or shape evolution is 
apparent in the more sparsely sampled 2014 LCOGT data.}.
Through these we could also determine the individual
masses $M_{\rm NS}$ and $M_c$, the companion's surface gravity
$\log g$, as well as other related parameters such as the irradiation
temperature $T_{\rm irr}$ (defined as $(T_{\rm day}^4-T_{\rm
  night}^4)^{1/4}$),  the irradiation efficiency $\eta$ ($4\pi a^2\sigma
T_{\rm irr}^4/\dot E$), and the volume-averaged Roche-lobe filling
factor $R/R_L$.  We assumed corotation, although
for a binary period of 15\,hr the companion may not be tidally
locked.  We also assumed a gravity darkening coefficient $\beta=0.08$
(with effective temperature $\propto g^\beta$ corresponding to a fully convective envelope;
\citealt{Lucy:67:GravityDarkening}).  We used 192 individual walkers to
explore the parameter space and followed each for 20,000 iterations.
We then rejected the first 200 iterations as ``burn-in'', and only
included every 173rd point (based on observed autocorrelation lengths
of 80--180 iterations for individual parameters).  The results of the
MCMC are shown in Figure~\ref{fig:fit} and given in
Table~\ref{tab:fit}, where we list the mean of the posterior pdfs
along with 68\% confidence limits on individual parameters.  Most of
the two-dimensional marginalized pdfs are relatively well determined.
The distance is a factor of 2 higher than the DM distance, although this is
not uncommon for pulsars out of the Galactic
Plane \citep{Roberts:11:BWandRedbacks,Kaplan:13:J1816} 
We do not have a strong constraint on the
inclination other than $\cos i<0.38$ at 95\% confidence
($i>68\degr$).  Even though the uncertainties on $T_{\rm night}$ and
$T_{\rm day}$ overlap, there is actually a well-determined irradiation
efficiency as $T_{\rm night}$ and $T_{\rm day}$ are highly
correlated.  The best-fit lightcurve is given in Figure~\ref{fig:lightcurve}.

To explore the effects of our assumptions for corotation and gravity
darkening, we did a limited series of experiments.  We tried a
corotation factor of 0.5 (rotation at half of the orbital speed) and a
gravity darkening factor $\beta=0.04$.  In both cases the majority of
the fitted parameters remain within 1-$\sigma$ of their nominal
values.  Only the filling factor changes slightly (still by
$<2\sigma$), and this is largely to keep the shape of the photometric
variations similar. 

\begin{deluxetable}{l r c l}
\tablecaption{Best-fit Binary Parameter Values for \psr\ Based on the
  MCMC \texttt{Icarus} Analysis\label{tab:fit}}
\tablehead{
\colhead{Parameter} & \multicolumn{3}{c}{Value}
}
\startdata
\cutinhead{Fitted Parameters}
$\cos i$\tablenotemark{a} & 0.16 & $\pm$ & 0.12 \\
$T_{\rm night}$\tablenotemark{b} (K) & 5094 & $\pm$ &90\\
$T_{\rm day}$\tablenotemark{c} (K) & 5124 & $\pm$ & 92 \\
$K$\tablenotemark{d} (km\,s$^{-1}$) & 250.3 & $\pm$ & 4.3 \\
dm\tablenotemark{e} & 11.3 &$\pm$& 0.1\\
$A_V$\tablenotemark{f} (mag) & 0.14 & $\pm$ & 0.09 \\
$f$\tablenotemark{g} & 0.82 & $\pm$ & 0.03 \\
$\chi^2$ / DOF & 558.9 & / & 328\\
\cutinhead{Derived Parameters}
$d$\tablenotemark{h} (pc) & 1833 & $\pm$ & 110 \\
$\log(g)$\tablenotemark{i} & 4.06 &$\pm$& 0.01\\
$q$\tablenotemark{j} & 3.94 &$\pm$& 0.07 \\
$M_{\rm c}$\tablenotemark{k} ($M_\odot$) & 0.44 &$\pm$&0.04\\
$M_{\rm NS}$\tablenotemark{l} ($M_\odot$) & 1.74 &$\pm$&0.18\\
$i$\tablenotemark{m} (deg) & 80.5 &$\pm$&7.0\\
$T_{\rm irr}$\tablenotemark{n} (K) & 2001 & $\pm$ & 91 \\
$\eta$\tablenotemark{o} (\%) & 3.0 & $\pm$& 0.6 \\
$R/R_L$\tablenotemark{p} & 0.95  & $\pm$ & 0.01 
\enddata
\tablenotetext{a}{Cosine of the inclination angle $i$.}
\tablenotetext{b}{Night-side temperature of the companion (i.e., the
  side facing away from the pulsar.}
\tablenotetext{c}{Day-side temperature of the companion (i.e., the
  side facing toward the pulsar.}
\tablenotetext{d}{Radial velocity amplitude of the companion.}
\tablenotetext{e}{Distance modulus.}
\tablenotetext{f}{$V$-band extinction.}
\tablenotetext{g}{Ratio of the radius of the companion pointing
  to the pulsar to that of the L1 point.}
\tablenotetext{h}{Distance.}
\tablenotetext{i}{Volume-averaged surface gravity of the companion.}
\tablenotetext{j}{Mass ratio.}
\tablenotetext{k}{Companion mass.}
\tablenotetext{l}{Pulsar mass.}
\tablenotetext{m}{Inclination angle.}
\tablenotetext{n}{Irradiation temperature, $(T_{\rm day}^4-T_{\rm
    night}^4)^{1/4}$.}
\tablenotetext{o}{Irradiation efficiency, $4\pi a^2 \sigma T_{\rm
    irr}^4/\dot E$, where $a$ is the semi-major axis.}
\tablenotetext{p}{Ratio of the volume-averaged radius to the 
volume-averaged Roche lobe radius.}
\end{deluxetable}

\section{Long-term photometry} \label{sec:longterm}

In Figure \ref{fig:longterm_byyear} we show the LINEAR, CRTS, PTF, P60
$r^\prime$, and LCOGT $r^\prime$
photometry from 2003--2014.  A secular evolution towards lower companion
brightness is
apparent in all surveys.  We examined photometry from nearby
stars of comparable brightness and found no long-term or phase-dependent 
trends.

To quantify this, we fit the data with an analytic model:
\begin{multline}
m(\phi) = m_0 + a_1\cos 2\pi \phi + a_2 \cos 4 \pi \phi + a_3 \cos
6\pi \phi \\ + a_4\sin 2\pi \phi + a_5\sin 4\pi \phi
\label{eqn:lc_fit}
\end{multline}
where $m(\phi)$ is the magnitude at orbital phase $\phi$ and $\phi$ is
in the range $0\rightarrow 1$.  
The coefficient $a_1$ is dominated by irradiation, $a_2$ is
dominated by ellipsoidal modulation, and $a_3$
accounts for residual distortion.  Of the sine coefficients, $a_4$ captures
relativistic beaming and $a_5$ fits residual distortions.

We fit for each coefficient
$m_0,a_1,a_2,a_3,a_4,a_5$ in each time bin in
Figure~\ref{fig:longterm_byyear} after normalizing the values of $m_0$ to
the PTF $R$-band.  Parameter agreement between the different instruments is
acceptable, as indicated by the reasonable goodness of fit ($\chi^2/{\rm
dof} = 1004.5/981$).
Table \ref{tab:fit_longterm} lists the best fit parameters.

\begin{table*}
\begin{center}
\begin{tabular}{|l|cccccc|ccc|}
\hline
Years & $m_0$ & $a_1$ & $a_2$ & $a_3$ & $a_4$ & $a_5$ & $\chi^2$ & dof &
$\chi^2_\nu$\\
\hline
2003-2004 & 16.545\ppm0.003 & 0.004\ppm0.005 & 0.114\ppm0.004 &
-0.010\ppm0.005 & 0.003\ppm0.005 & 0.018\ppm0.005 & 144.0 & 113 & 1.27\\
2005-2006 & 16.573\ppm0.003 & 0.010\ppm0.004 & 0.113\ppm0.005 &
-0.013\ppm0.004 & -0.006\ppm0.005 & 0.012\ppm0.004 & 177.5 & 205 & 0.87\\
2007-2008 & 16.598\ppm0.003 & -0.001\ppm0.004 & 0.117\ppm0.004 &
0.001\ppm0.005 & 0.013\ppm0.005 & 0.021\ppm0.005 & 211.7 & 184 & 1.15\\
2009-2010 & 16.637\ppm0.002 & -0.010\ppm0.002 & 0.117\ppm0.002 &
-0.014\ppm0.002 & -0.004\ppm0.002 & 0.014\ppm0.002 & 228.0 & 220 & 1.04\\
2011-2012 & 16.644\ppm0.003 & -0.004\ppm0.004 & 0.117\ppm0.003 &
-0.017\ppm0.003 & -0.022\ppm0.003 & 0.015\ppm0.004 & 127.4 & 142 & 0.90\\
2013-2014 & 16.680\ppm0.002 & -0.003\ppm0.002 & 0.106\ppm0.002 &
-0.005\ppm0.003 & -0.022\ppm0.002 & 0.015\ppm0.002 & 116.0 & 117 & 0.99\\
\hline
Total & &&&&&& 1004.5 & 981 & 1.02 \\
\hline
\end{tabular}
\caption{Best-fit coefficients (Equation \ref{eqn:lc_fit}) for the photometric evolution of
\psr by year.\label{tab:fit_longterm}}
\end{center}
\end{table*}

\begin{figure}
\plotone{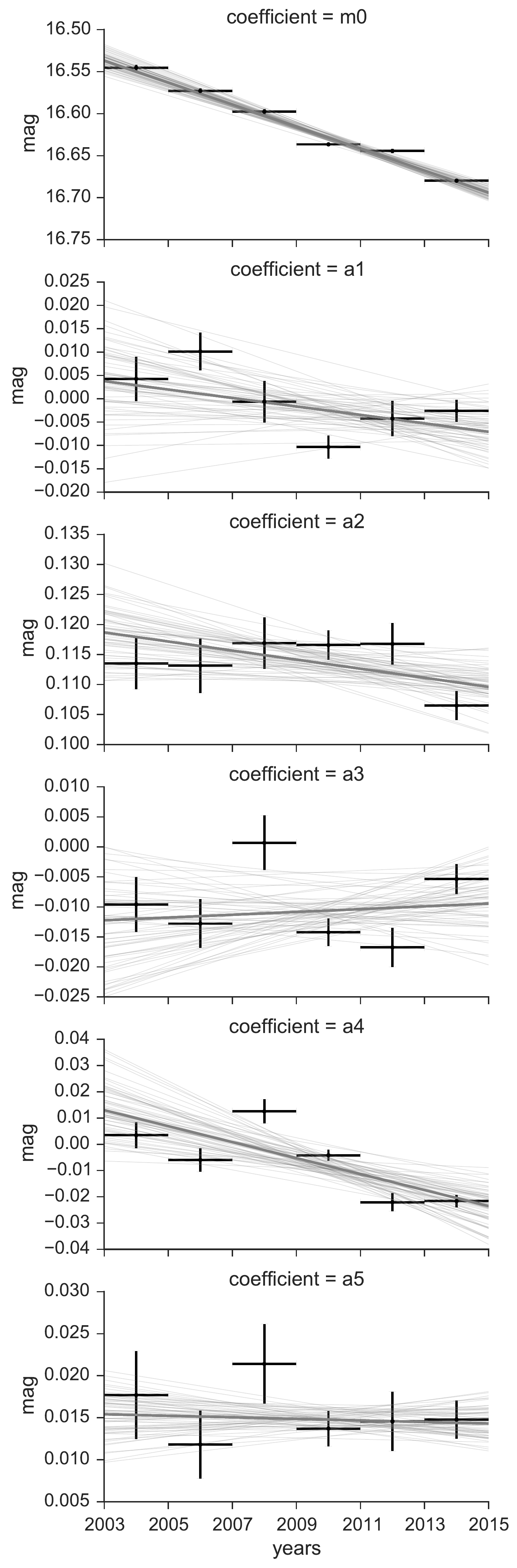}
\caption{Temporal evolution of the best-fit lightcurve parameters (Equation
\ref{eqn:lc_fit}).  
The dark gray lines are the best fit linear model; random deviates drawn
from the fit covariance matrix are plotted in light gray.  
\label{fig:parameter_evolution}}
\end{figure}

Figure \ref{fig:parameter_evolution} shows the time evolution of the best
fit parameters.  The most striking change is in the evolution of $m_0$,
which 
becomes systematically larger with time (i.e., the
companion becomes fainter).  
The best-fit linear change is $\dot
m_0=0.0131\pm0.0010\,{\rm mag\,yr}^{-1}$.  

We tested for possible evolution in the orbital period to ensure that
phase mis-specification was not the source of the observed decline in $m_0$. 
(The integrated statistical uncertainty from the radio ephemeris 
on the orbital phase is just 0.16\,sec over our ten-year baseline.)
The best fit to our data
is consistent with no change in the orbital period; the $1\sigma$
confidence limits are $-2.1\times10^{-10} < \dot{P}_{\rm orb} <
2.8\times10^{-9}$.  Additionally, if we discard all phase information and
simply compute the median magnitudes in two-year bins, we see a comparable
decline.  We conclude that the observed dimming is robust to variations of
the orbital period at this scale 
and assume $\dot{P}_{\rm orb} = 0$ for the remaining discussion.

Despite the clear linear decline in $m_0$, 
secular changes in the other coefficients are less
apparent.  Our ability to discern shape changes is hindered by the
heterogeneous dataset, with instruments, phase coverage, and precision
varying from year to year.  
Few of the
linear fits to the parameter evolution shown in Figure
\ref{fig:parameter_evolution} are formally acceptable; 
$\chi^2$ values for $m_0$--$a_5$ are
45.9, 17.5, 8.0, 16.6, 25.8, and 3.0, all for 4 degrees of freedom.
All of the coefficients except $a_4$ appear generally consistent with a
constant value, but with one or two stochastically outlying points that we
attribute to the mixed dataset.

The behavior the coefficient $a_4$ is particularly difficult to
interpret.  This sine term corresponds to relativistic beaming.  For our
fit radial velocity amplitude $K = 250$\,km\,s$^{-1}$, the expected
amplitude is $\approx 4$\,mmag \citep{Loeb:03:Beaming}.  However, our fit
values are nearly an order of magnitude larger and change sign, implying an
physically impossible reversal of the orbital velocity.  Moreover, the sine
coefficient $a_5$ is inconsistent with the value of zero we would expect
from a self-consistently modeled lightcurve \citep{Mazeh:10:CoRoTBeaming}.

We also fit Equation \ref{eqn:lc_fit} to our best-fit binary model  from
\S~\ref{sec:binary}.  In this case $a_1$--$a_3$ took values very similar to
those obtained from direct fits to the data, but $a_4$ and $a_5$ were
approximately zero.  This suggests that our fit values of $a_4$ and $a_5$
are modeling (possibly spurious) features of the data not present in the
geometric approximation of the photosphere assumed by \texttt{Icarus}.
These apparent shape variations are most likely due to the complexities of
combining photometry with irregular phase coverage from several distinct
surveys.
More speculatively, however, it is possible that we are seeing an
additional emission component from outside the companion's Roche
lobe--for example, from the intrabinary shock region, if the
radio-eclipsing material there is not optically thin.  
\citet{Roberts:15:RedbacksXrays}
report that the X-ray emission from \psr shows an unusual
phase dependence, with two strong but asymmetric 
peaks bracketing the companion superior
conjunction, near the quadrature phases.  
Such a signal could put power into the $a_4$ and $a_5$ sine coefficients.
On the other hand, our optical spectroscopy of the
system does not display any features obviously distinct from a stellar
companion.

We explored potential physical origins of the observed long-term variation
of $m_0$.
For the companion to get
fainter, we could have a change in the radius (becomes smaller), in the
effective temperature (becomes cooler), or both.  
These could be due to perturbation by a past episode of active
accretion that occurred before 2003, when our photometry begins.
We tested these
possibilities by modifying our best-fit binary models.  

First, we changed
the Roche filling factor (equivalent to changing the radius).  As expected, as
the filling factor decreases the companion gets fainter, with a fractional
decrease of 0.01\%/yr required to match the change in $m_0$.  However, the
other coefficients also changed (indeed, we would expect the ellipsoidal
modulations to decrease as the filling factor decreases).  The most
significant change is in $a_2$, which would change by $-0.003\,{\rm
mag\,yr}^{-1}$.  As seen in Figure \ref{fig:parameter_evolution}, 
$a_2$ is essentially constant except for the final bin. The best-fit line
gives $\dot{a}_2 = -0.0007\pm0.0005\,{\rm mag\,yr}^{-1}$, smaller than
predicted under the assumption that the flux decline is due to the
companion shrinking.

The other simple option would be a change in effective temperature.
We decreased both the night-side (base) temperature and day-side
temperatures, with the irradiation temperature fixed. This may be
overly simplified, if the pulsar's irradiation is changing, but our fits
suggest the irradiation term $a_1$ is essentially constant.  In this
model the change in $m_0$ requires a decrease of about $10\,{\rm
  K\,yr}^{-1}$ in $T_{\rm night}$, which also leads to a change in $a_2$
 of $0.0001\,{\rm mag\,yr}^{-1}$, only slightly smaller than the observed value.
There is also a decrease in the mean $g-r$ color of $0.004\,{\rm
  mag\,yr}^{-1}$.  
Since we only have a single epoch of well-sampled
multi-band photometry we cannot look for color variations at this time. 

While a decreasing effective temperature is most consistent with our
current data, the thermal timescale of the star is 10\,Myr, 
much longer  than the yearly dimming we observe.  Even when limited to the
convective envelope, the thermal timescale is 3\,Myr.

Despite the clear trend towards dimmer companion magnitudes seen in CRTS,
LCOGT, and PTF, we are thus unable to form a self-consistent picture for
the physical cause of the variation.  Additional multicolor and high-SNR
monitoring in the years ahead should help resolve this ambiguity.

\section{Discussion} \label{sec:discussion}

While \psr clearly meets the observational criteria for identification as a
redback system---it is an eclipsing MSP with a nondegenerate companion of a
few tenths of a solar mass---its extreme parameters relative to other
systems in that population as well as its unusual position in evolutionary
phase space make it unique among currently known redbacks.

Our best-fit mass for the non-degenerate 
companion of \psr is $M_{\rm c} = 0.44\pm0.04\msun$,
one of the heaviest of known redback
systems.  (PSR\,J1723$-$2837 may have a heavier companion, but its mass is
poorly constrained \citep[0.4--0.7\msun;][]{Crawford:13:J1723Mass}.)
Combined with its long pulse period
\citep[7.6\,msec;][]{Hessels:11:GBTFermiSurvey} and strong magnetic field 
($B\sim1.6\times10^9$\,G; \citealp{Roberts:15:RedbacksXrays})
relative to
other redback systems \citep{Roberts:13:Spiders}, these data suggest that
the \psr system is early in its recycling phase.  

The pulsar companion is 95\% Roche-lobe filling, consistent with a
``quasi-Roche lobe overflow'' (qRLOF) state where irradiation feedback
during a previous accretion phase causes the companion to transfer more mass
than it would without irradiation, leaving it just below Roche filling
after the cessation of accretion \citep{Benvenuto:15:RedbackQuasiRLOF}.  As
discussed in that paper, however, the qRLOF model cannot account for the
$\sim$year timescale variations seen here as well as in the redback systems
that have been observed to transition between accreting and detached
states.  While accretion disk instabilities could account for the state
changes in the transitioning systems, they are unlikely to be relevant for
\psr.

The orbital period and companion mass of the \psr system 
appear compatible with standard
binary evolution of a neutron star and a normal main-sequence companion
under some initial conditions 
\citep[Figure 2 of][]{Podsiadlowski:02:LMXBEvolution}. 
Interestingly, the bifurcation of evolutionary
tracks into short and long period systems occurs near the position of \psr
in the $P_{\rm orb}$--$M_{\rm c}$ plane.

\begin{figure*}
\plotone{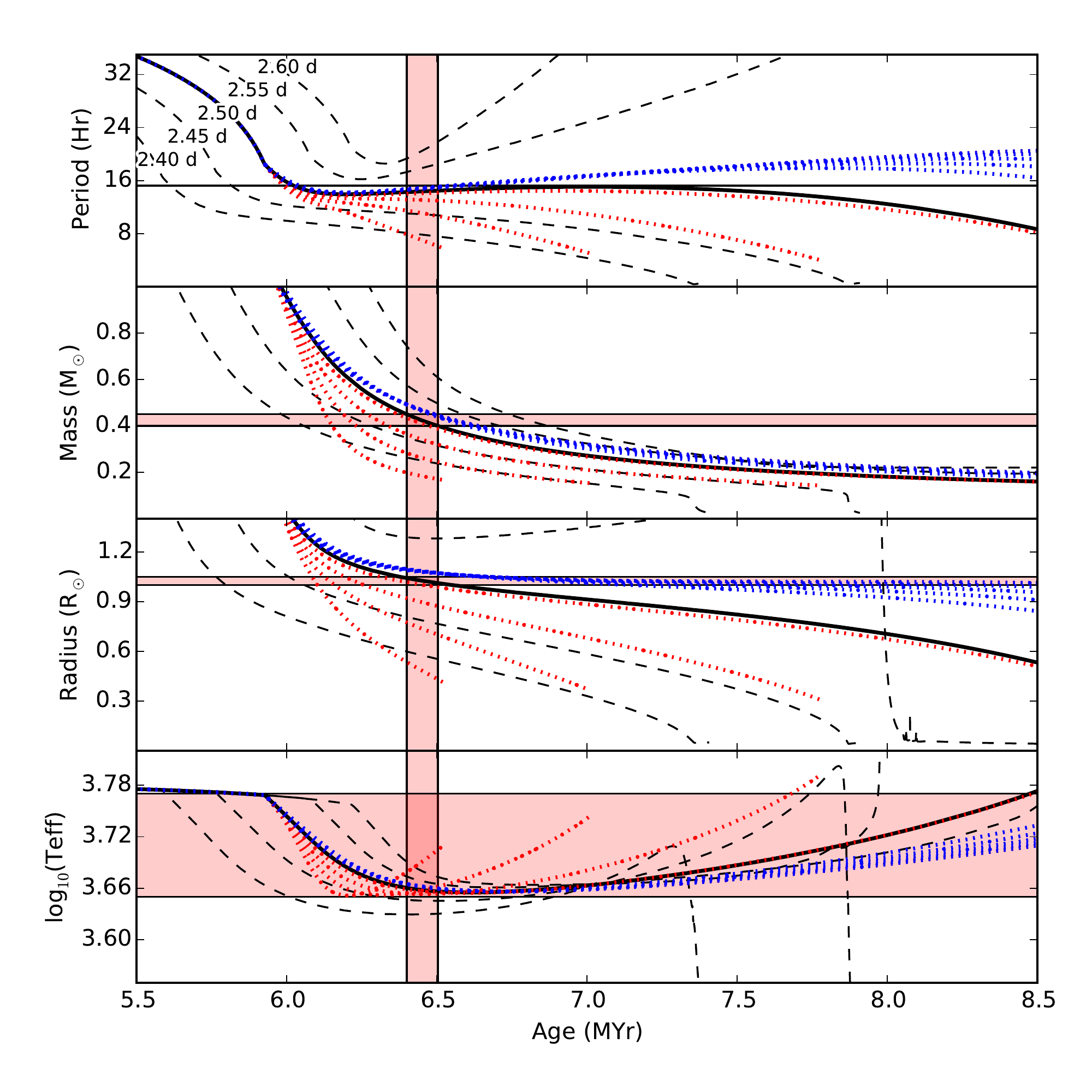}
\caption{
The evolutionary path of the orbital period, donor mass, donor radius, and
donor temperature for a
1.143\,$M_\odot$ zero-age main sequence donor orbiting a 1.6\,$M_\odot$
neutron star. The solid lines display the effect of varying the initial
orbital period in a range from 2.40 to 2.60\,d.
The dotted lines show the effects of varying mass loss assuming the 2.50\,d
initial period.
The blue dotted lines display the effect of mass loss at the companion,
with $\alpha = 0$ and $\beta = {0, 0.25, 0.5, 0.75, 1}$, from top to
bottom. The red dotted lines display the effect of mass loss at the neutron
star, with $\beta = 0$ and $\alpha = {0, 0.25, 0.5, 0.75, 1}$, from top to
bottom. Note that the uppermost blue and red lines (e.g. $\alpha, \beta =
0$) overlap.
On each plot, the horizontal red shaded region represents the 68\% probability
of the given parameter as inferred from our optical light curve modeling. The
vertical red shaded region corresponds to the age compatible with the companion
mass for the 2.50\,d model. 
\label{fig:binary_evolution}}
\end{figure*}

To explore the possible evolutionary history of \psr, we conducted a series
of simulations using the Modules for Experiments in Stellar Astrophysics
(MESA) code \citep{Paxton:11:MESA1,Paxton:13:MESA2,tmp_Paxton:15:MESA3}.
Figure \ref{fig:binary_evolution} shows the evolutionary path of a
1.143\,$M_\odot$ zero-age main sequence donor orbiting a 1.6\,$M_\odot$
neutron star. We examined the effect of varying the initial
orbital period in a range from 2.40 to 2.60\,d. For these curves we assumed
mass loss fractions in the vicinity of the donor and of the neutron star of
$\alpha = 0.2$ and $\beta = 0.5$, respectively, thus resulting in an
overall accretion efficiency by the neutron star $\epsilon = 0.2$. Effects
of evaporation through irradiation were neglected for reasons that will
become apparent later. Systems with initial orbital periods $\lesssim
2.50$\,d are converging \citep[case A
evolution;][]{Tauris:06:CompactBinaryEvolution} and will evolve to become
black widows and redbacks. At larger initial orbital separations, mass
transfer only sets in on the red giant branch \citep[case B
evolution;][]{Tauris:06:CompactBinaryEvolution} and so these systems will
diverge to become pulsar binaries with He-core white dwarf companions. 
At 2.50\,d the
\psr system is marginally diverging, but angular momentum loss due to
gravitational wave radiation eventually makes the orbit shrink once the
main phase of mass transfer has stopped. Also displayed are evolutionary
tracks for 2.50\,d initial periods having different mass-loss parameters.

Given our observational 
constraints on companion mass, radius, and temperature as well as
the orbital period, we can determine which evolutionary tracks are feasible
by requiring that they produce consistent observables at a single age.
As we can see, our evolution tracks indicate that the
initial orbital period must be very close to the bifurcation period and
require $\alpha \sim 0.25$, while the effect of $\beta$ is negligible in
the considered range of evolution.  Changing the assumed initial mass ratio
qualitatively only changes the timescale of the evolution.

For a reasonable range of initial neutron star masses, the evolution at a
similar mass ratio remains unchanged, except for a rescaling of the initial
orbital period. In the $M_{\rm NS, i} = 1.6\,M_\odot$ scenario, the accreted
mass is $\sim 0.2\,M_\odot$ which implies a neutron star mass of
$1.8\,M_\odot$
which is similar to the value inferred from our light curve modeling. While we
have not searched for compatible solutions for an exhaustive range for
parameters at other mass ratios, we found that the mass ratio needs to be close
to the presented one if $\alpha = 0.2$ and $\beta = 0.5$. In every case, we
found that the only systems able to reproduce the observed properties are
always close to the bifurcation period.

In light of our binary evolution work, we conclude that \psr is in a
very specific region of the phase space. At present the system is close to
Roche-lobe filling and thus given its mass could not have been in a
significantly tighter orbit in the past. 
Irradiation from the pulsar is very mild because of the large orbital
separation and, despite the fact that it could have been larger in the past when
the pulsar was spinning faster, it is unlikely that it has ever been very
effective. Indeed, as demonstrated by
\citet{Chen:13:RBBWEvolution}, evaporation through
irradiation leads to an increase of the orbital period and so this mechanism
could not have happened significantly given that the current orbit is close to
the shortest separation that it has ever been. This also justifies our
assumption to neglect the effect of irradiation.

The fate of \psr is somewhat ambiguous as it lies right at the
boundary between systems that will become redbacks/black widows and those that
will widen and form a He-core white dwarf. Gravitational wave radiation should
cause the system to shrink, but whether the orbit may become significantly compact
within the next billion years or so depends  on how further mass transfer and
irradiation will operate. As it is, the orbit might well still be marginally
widening if residual mass transfer is still operating.

Our modeling shows that it is possible to  produce the observed
high pulsar mass ($M_{\rm NS} = 1.74\pm0.2\msun$) with some assumptions
under standard recycling scenarios.   
Other evolutionary scenarios are not ruled out, however.
Production of a NS with this mass via Accretion Induced Collapse rather
than recycling appears possible, but would require a rapidly-rotating WD
and high accretion efficiency \citep{Smedley:15:RedbackAIC}.
Another possibility is that the neutron star was born heavy, with a mass
near the observed 1.74\msun
\citep{Tauris:11:NSBornHeavy}.

The origin of the 13\,mmag\,yr$^{-1}$ 
dimming of the companion of \psr discovered in
this work remains uncertain.
If the companion is shrinking, we would expect to see larger changes in the
ellipsoidal modulation of the phased lightcurve than are observed.  Cooling
of the companion is compatible with these data, but the expected thermal
timescales for the system are much longer than the observed yearly
variations.  
Future multi-color observations will be
valuable in monitoring the ongoing evolution of this unique system.
Further investigation may reveal if \psr will eventually destroy its
companion or if---like the katipo spider found in New Zealand---it is a
redback that shuns cannibalism.

\acknowledgments

T.A.P, E.C.B., and S.T. acknowledge partial support through NASA Grant No.
NNX12AO76G.  R.P.B. has received funding from the European Union Seventh
Framework Programme under grant agreement PIIF-GA-2012-332393.  J.W.T.H.
acknowledges funding from an NWO Vidi fellowship and ERC Starting Grant
``DRAGNET'' (337062).  E.C.B. and T.A.P. thank the Aspen Center for Physics
and the NSF Grant \#1066293 for hospitality during the editing of this
paper.  

This paper is based in part on observations obtained with the Palomar 48-inch
Oschin telescope and the robotic Palomar 60-inch telescope at the Palomar
Observatory as part of the Palomar Transient Factory project, a scientific
collaboration among the California Institute of Technology, Columbia
University, Las Cumbres Observatory, the Lawrence Berkeley National
Laboratory, the National Energy Research Scientific Computing Center, the
University of Oxford, and the Weizmann Institute of Science; and the
Intermediate Palomar Transient Factory project, a scientific collaboration
among the California Institute of Technology, Los Alamos National
Laboratory, the University of Wisconsin, Milwaukee, the Oskar Klein Center,
the Weizmann Institute of Science, the TANGO Program of the University
System of Taiwan, and the Kavli Institute for the Physics and Mathematics
of the Universe. 

The CSS survey is funded by the National Aeronautics and Space
Administration under Grant No.\,NNG05GF22G issued through the Science
Mission Directorate Near-Earth Objects Observations Program.  The CRTS
survey is supported by the U.S.~National Science Foundation under
grants AST-0909182, AST-1313422, and AST-1413600.

This work makes use of observations from the Las Cumbres Observatory Global
Telescope Network.

{\it Facilities:} \facility{Hale (Double Beam Spectrograph)},
\facility{PO:1.2m (Palomar Transient Factory)}, \facility{PO:1.5m},
\facility{LINEAR}, \facility{CSS}, \facility{LCOGT}

\bibliographystyle{yahapj}
\bibliography{ptf}

\begin{turnpage}
\begin{figure*}
\includegraphics[width=0.8\textheight]{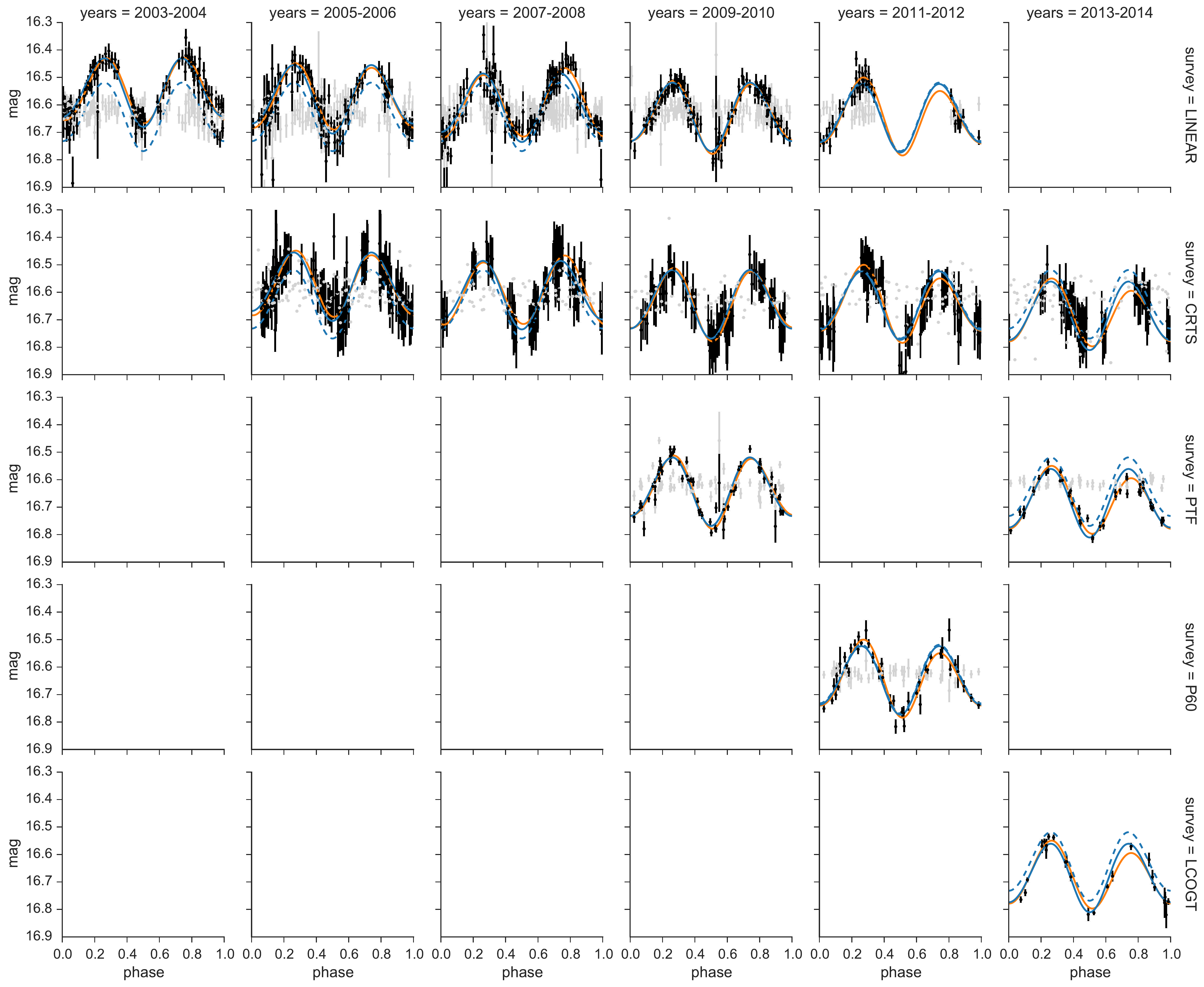}
\caption{Long-term photometry for \psr from LINEAR, CRTS, PTF, P60
($r^\prime$), and LCOGT ($r^\prime$).  Data
are grouped in two-year intervals.  Orbital phases are computed from the
radio ephemeris.
Photometry of \psr is shown in black, while that of the nearby comparison
star is shown in grey.  Errors on the CRTS companion photometry are omitted for
clarity but are comparable to that of the pulsar.
We fit and removed constant offsets from LINEAR ($-0.82$\,mag), CRTS
($-0.08$\,mag), P60 ($-0.11$\,mag), and LCOGT ($-0.11$\,mag) to match PTF.
Model fits by year with fixed sine and cosine 
amplitudes are shown in solid blue lines, while the fit to the
2009--2010 data is replicated in blue dashed lines to illustrate the temporal
evolution of the source.  Model fits by year with the sine and cosine amplitudes free to vary are plotted in orange.
\label{fig:longterm_byyear}}
\end{figure*}
\end{turnpage}
\clearpage

\end{document}